\documentclass[aps,prd,amsfonts,amsmath,superscriptaddress,twocolumn,nofootinbib]{revtex4}

\usepackage{color} 
\usepackage{graphicx} 
\usepackage{epstopdf}

\newcommand {\be} {\begin{equation}} 
\newcommand {\ba}{\begin{eqnarray}} 
\newcommand {\ee} {\end{equation}} 
\newcommand{\ea} {\end{eqnarray}}

\begin{document}

\title{Proton structure corrections to hyperfine splitting in muonic
hydrogen}

\author{Carl E.\ Carlson}
\affiliation{Helmholtz Institut Mainz, Johannes Gutenberg-Universit\"at, D-55099 Mainz, Germany}
\affiliation{Department of Physics, College of William and Mary, Williamsburg, VA 23187, USA}

\author{Vahagn Nazaryan}
\affiliation{Center for Advanced Medical Instrumentation, Department of Physics, Hampton University, Hampton, VA 23668}

\author{Keith Griffioen}
\affiliation{Department of Physics, College of William and Mary, Williamsburg, VA 23187, USA}

\date{January 17, 2011}

\begin{abstract}

We present the derivation of the formulas for the proton structure-dependent terms in the hyperfine splitting of muonic hydrogen. We use compatible conventions throughout the calculations to derive a consistent set of formulas that reconcile differences between our results and some specific terms in earlier work.  Convention conversion corrections are explicitly presented, which reduce the calculated hyperfine splitting by about $46$ ppm. We also note that using only modern fits to the proton elastic form factors gives a smaller than historical spread of Zemach radii and leads to a reduced uncertainty in the hyperfine splitting.  Additionally, hyperfine splittings have an impact on the muonic hydrogen Lamb shift/proton radius measurement, however the correction we advocate has a small effect there.

\end{abstract}

\maketitle

%%%%%%%%%%%%%%%%%%%%%%%%%%%%%%%%%%%%%%%%%%%%%%%%

\section{Introduction}

The recent conundrum with the proton charge radius~\cite{Pohl:2010zz} inspires reconsideration of the corrections that enter into determinations of proton size dependent energy splittings in hydrogen atoms, particularly muonic hydrogen.  This note is about the determination of the polarizability and elastic corrections to the hyperfine splitting in muonic hydrogen, where there is a disagreement implicit in the literature which should be resolved.  The numerical consequences of the resolution have a small effect upon current measurements, but could be consequential when accurate measurements of muonic hydrogen hyperfine splitting become available and if other sources of uncertainty are reduced.

For the leading order proton structure-dependent part of the hyperfine splitting (HFS) in muonic hydrogen, the calculation as commonly implemented requires, we claim,  further correction due to overlapping terms between the elastic and polarizability parts of the calculation.  The situation occurs because terms can be moved from the inelastic to the elastic part of the calculation using the Gerasimov-Drell-Hearn-Hosada-Yamamoto sum rule (which relates the proton anomalous moment to inelastic cross sections)~\cite{Gerasimov:1965et,Drell:1966jv,hosoda1966}, and details of the move are handled differently in different sources.  For cases known to us, the differences are of no numerical consequence for ordinary hydrogen, but are noticeable for muonic hydrogen.  Articles on this subject often calculate just the elastic~\cite{Bodwin:1987mj,Pachucki:1996zza} or the just inelastic corrections~\cite{Faustov:2001pn}, and sometimes both in the same article~\cite{Nazaryan:2005zc,Carlson:2008ke}.  A calculation of the proton structure corrections that combines elastic and inelastic (under the heading of ``polarizability'') results from incompatible sources will obtain a deficient total, and such calculations appear to exist in practice.  At a practical level, given that many full muonic HFS articles (\textit{e.g.,}~\cite{Dupays:2003zz,Martynenko:2004bt}) also usefully compile higher-order corrections and small terms from other effects, perhaps the best procedure is not to start from scratch, but to add to the existing calculations the needed term that coordinates the elastic plus inelastic total.

Part of the current motivation for discussing the HFS is the conundrum occasioned by the recent measurement of the proton charge radius using the Lamb shift in muonic hydrogen~\cite{Pohl:2010zz}.  The HFS has some, albeit small, effect there.  The Lamb shift is the energy difference, mostly due to QED effects, between the $2S_{1/2}$ and $2P_{1/2}$ hydrogen levels.  More precisely, given that each of the levels is hyperfine split, it is the energy difference between the properly weighted average of the hyperfine levels of the $2S_{1/2}$ and $2P_{1/2}$ states.  The goal is to use the measured Lamb shift to infer the extra overall shift of the $2S_{1/2}$ energy due to proton size corrections.  The experiment actually measures the energy difference between the $2S_{1/2}^{F=1}$ state (where $F$ is the total lepton + proton angular momentum) and the $2P_{3/2}^{F=2}$ states, and relies upon calculation of the $2S_{1/2}$ HFS and the $2P_{3/2}$--$2P_{1/2}$ splitting to obtain the Lamb shift.  The $P$-states involve no proton structure corrections to the accuracy required.  The $2S_{1/2}^{F=1}$ state is shifted up by (1/4) of the full HFS, so a change in the HFS has an effect upon the Lamb shift extraction.  However for the correction discussed in this paper, the effect translates into less than 1\% of the current discrepancy between the muonic hydrogen value of the charge radius~\cite{Pohl:2010zz} and the CODATA value~\cite{Mohr:2008fa}, which is based mainly on the ordinary hydrogen Lamb shift and confirmed by electron scattering data~\cite{Sick:2003gm}.

Our outline for this paper is as follows.  Many of the formulas we use were given in our earlier work~\cite{Nazaryan:2005zc,Carlson:2008ke}, but without derivations being shown.  Not much detail was given as well for the muonic polarizability calculation in~\cite{Faustov:2001pn}.  Since there are some differences to be assessed, we show the calculation in moderate detail in Sec.~\ref{sec:formulas}.  It parallels the well-known calculation for negligible lepton mass~\cite{Iddings:1965zz,Drell:1966kk,DeRafael:1971mc}, but now keeps the mass general.  A reader who does not wish to check the details on a first reading can proceed directly to Sec.~\ref{sec:correction}, where we isolate the term that should be added to some existing muonic HFS calculations, evaluate it numerically, and compare its size to other known terms. The effect of this change to the HFS upon the current measurement of the proton charge radius from the Lamb shift is shown in Sec.~\ref{sec:lamb}.  A short discussion is offered in Sec.~\ref{sec:end}

%%%%%%%%%%%%%%%%%%%%%%%%%%%%%%%%%%%%%%%%%%%%%%%%

%%%%%%%%%% 
\begin{figure}[b] 
\begin{center}

\includegraphics[width = 6.0 cm]{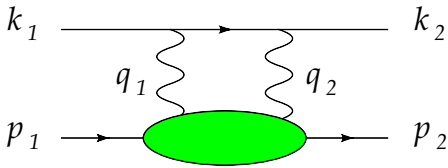}

\caption{Diagram giving proton structure-dependent corrections to the
hyperfine splitting.} 
\label{fig:hfs_box} 
\end{center} 
\end{figure}
%%%%%%%%%%

%%%%%%%%%%%%%%%%%%%%%%%%%%%%%%%%%%%%%%%%%%%%

\section{Hyperfine splitting}

\label{sec:formulas}

%%%%%%%%%%%%%%%%%%%%%%%%%%%%%%%%%%%%%%%%%%%%

The relevant diagram is shown in Fig.~\ref{fig:hfs_box}.

We neglect the Fermi momentum of the atomic lepton, and note that the proton leg plus photons is the same as off-shell forward Compton scattering on the proton,
which is calculated from the matrix element of a time ordered product
\begin{eqnarray}
T_{\mu\nu}(q) &=& \frac{i}{2\pi m_p} \int d^4\xi \ e^{iq{\cdot}\xi}
\left\langle pS \right| {\cal T} j_\mu(\xi) j_\nu(0) \left| pS \right\rangle  , \quad  
\end{eqnarray}

\noindent   where $j_\mu$ is the electromagnetic current and the states are normalized by $\langle pS | p'S  \rangle = 2E (2\pi)^3 \delta^3(p-p')$ (for the same $S$).  We need only the spin
dependent part of $T_{\mu\nu}$, which is antisymmetric in its indices and is expanded using the scalar functions $H_1$ and $H_2$, 
\begin{eqnarray} 
T^A_{\mu\nu} &=& \frac{i}{m_p} \,
\epsilon_{\mu\nu\alpha\beta} q^\alpha \times 
\\	\nonumber
&\times&   \bigg[ S^\beta \, H_1(q_0,Q^2) + \frac{ p{\cdot}q \
S^\beta - S{\cdot}q \ p^\beta }{ m_p^2 } \, H_2(q_0,Q^2) 	\bigg] 
\,;
\end{eqnarray}

\noindent    $q_0 = q^0$ is given in the proton rest frame, $q_0 = p\cdot q/m_p$, and $S^\beta$ is the proton spin
vector.

From the definitions
\ba 
H_1(-q_0,Q^2) &=& H_1(q_0,Q^2)   \,, 
\nonumber \\ 
H_2(-q_0,Q^2) &=& - H_2(q_0,Q^2) \,.
\ea

The functions $H_{1,2} (q_0,Q^2)$ are not known either from \textit{ab initio} calculation or measurement.  However, the functions can be constructed from their imaginary parts.    The imaginary parts come from the contribution where the intermediate electron and hadron states, in Fig.~\ref{fig:hfs_box}, are on-shell physical states,  which in turn are squares of electron-proton elastic or inelastic scattering amplitudes.   Hence we can obtain $H_{1,2} (q_0,Q^2)$ using dispersion relations and data on electron-proton scattering cross sections.

In terms of standard notation, electron-proton scattering cross sections are given from the hadronic tensor
\begin{equation}
W_{\mu\nu}(q) = \frac{1}{4\pi } \int d^4\xi \ e^{iq{\cdot}\xi}
\left\langle pS \right| 
\left[  j_\mu(\xi) , j_\nu(0) \right]    \left| pS \right\rangle   , 
\end{equation}
and the spin-dependent part of $W_{\mu\nu}$ is 
\begin{eqnarray} 
W^A_{\mu\nu} &=& i \,
\epsilon_{\mu\nu\alpha\beta} q^\alpha \times 
\\	\nonumber
&\times&   \bigg[ S^\beta \, g_1(q_0,Q^2) + \frac{ p{\cdot}q \
S^\beta - S{\cdot}q \ p^\beta }{ m_p^2 } \, g_2(q_0,Q^2) 	\bigg] 
\,. 
\end{eqnarray}
Hence (for $q_0>0$),
\ba 
{\rm Im\,} H_1(q_0,Q^2) &=& \frac{1}{q_0} \  g_1(q_0,Q^2) \ , 
\nonumber	\\[1.25ex] 
{\rm Im\,} H_2(q_0,Q^2) &=&  \frac{m_p}{q_0^2} \   g_2(q_0,Q^2)   \ .
\ea
The spin-dependent structure functions
$g_1$ and $g_2$ are measured (over some kinematic range) in inelastic polarized $e$-$p$ scattering at laboratories including CERN, SLAC, HERMES, ELSA, and JLab.  For elastic scattering they are given in terms of the elastic form factors by
\ba
g_1^{elastic} &=& \frac{1}{2} F_1 G_M \, \delta(1-x)  \ ,
						\nonumber \\
g_2^{elastic} &=& - \frac{1}{2} \tau F_2 G_M \, \delta(1-x)  \ ,
\ea
where $x = Q^2/(2p \cdot q)$, $F_1(Q^2)$ and $F_2(Q^2)$ are the Dirac and Pauli form factors, respectively, and $G_M$ and (for completeness) $G_E$ are
\ba
G_M &=& F_1 + F_2 \,,   \nonumber \\
G_E &=& F_1 - \frac{Q^2}{4m_p^2} F_2   \,.
\ea

\begin{figure}

\medskip
\includegraphics[width=3.4in]{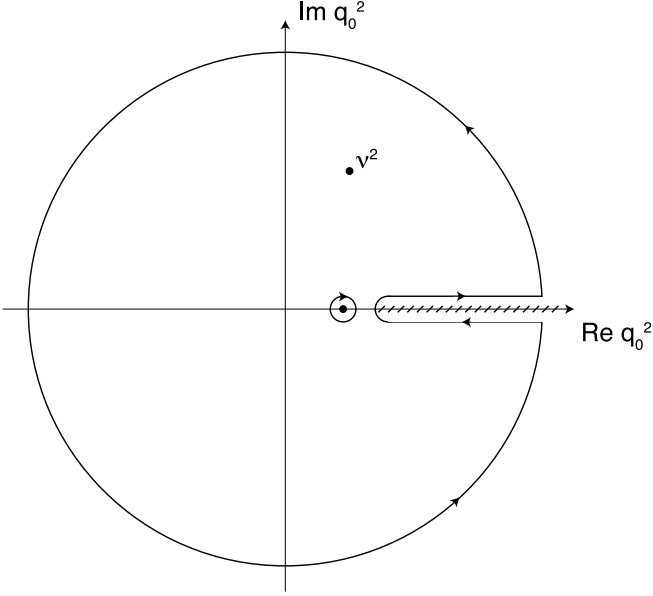}
\caption{Contour in complex $q_0^2$ plane for Cauchy integral.}
\label{fig:cauchy}

\end{figure}

The functions $H_{1,2}(q_0,Q^2)$ have poles in $|q_0|$ at the elastic point and cuts beginning at the inelastic threshold
\be
\nu_{th} = m_\pi + (m_\pi^2 + Q^2)/(2 m_p)
\ee
and going to infinity along the real axis.   We use the integration contour illustrated in the complex $q_0^2$ plane in Fig.~\ref{fig:cauchy}, assume zero contribution from the part of the contour at infinite distance, and obtain the dispersion relations

%%%%%%%%%%%%%%%%%%%%%%%
%%%%%%%%%%%%%%%%%%%%%%%

\ba 
H_1(q_0, Q^2)  &=& 
- \frac{2m_p}{\pi} 
	\frac{ q^2 F_1(Q^2) G_M(Q^2) } { (q^2 + i\epsilon)^2 - 4m_p^2
q_0^2 } 
\nonumber \\ 
&+& \frac{2}{\pi} \int_{\nu_{th}}^\infty  
\frac{d{\nu}}{{\nu}^2 - q_0 ^2} \, g_1(\nu, Q^2)  \,,
				\\[1ex]
H_2(q_0,Q^2) &=& - \frac{2m_p}{\pi} 
\frac{ m_p \nu F_2(Q^2) G_M(Q^2) } 
{ (q^2 + i\epsilon)^2 - 4m_p^2 q_0 ^2 } 
\nonumber \\ 
&+& \frac{2 m_p q_0}{\pi} 
\int_{\nu_{th}}^\infty  \frac{d\nu}{{\nu}^2 - q_0 ^2} \
\frac{1}{ {\nu}^2}  \  g_2(\nu,Q^2)  \,. \quad 
\ea

Evaluating the two-photon, structure-dependent, contributions to the HFS, 
Fig.~\ref{fig:hfs_box}, one obtains after making a Wick rotation,
\begin{align}
& \frac{ E^{box}_{2\gamma} }{ E_F } =
 \frac{ \alpha m_\ell } { (1+\kappa_p) \pi^2 } \int \frac{ d^4 Q}{ Q^2} 
 \frac{1}{ Q^4 + 4m_\ell^2 Q_0^2 }	\times 
					\\[1ex]	\nonumber
&\times		\bigg\{ ( 2 Q^2 + Q_0 ^2 ) H_1(i Q_0,Q^2)
		- \ \ 3 Q^2 Q_0 ^2 
\frac{H_2(i Q_0,Q^2)}{i m_p Q_0}  \bigg\} ,
\end{align}
where $E_F$ is the Fermi energy
\be
    E_F^p=\frac{8 \alpha^3 m_r^3 }{3\pi} \mu_B\mu_p
    	=	\frac{16  \alpha^2}{3} \frac{\mu_p}{\mu_B}  
		\frac{ R_\infty }{ \left( 1 + m_\ell/m_p \right)^3 } 	\,,
\ee
and $m_r$ is the lepton reduced mass.  The Wick rotation entails the notations
\ba 
q_0 &=& i Q_0  \,,
		\nonumber \\ 
d^4q &=&  i \, d Q_0\,  d^3q = i \, d^4 Q  \,,
		\nonumber	\\ 
q^2 &=& q_0 ^2 - \vec q^{\ 2} =  - Q^2	\,. 
\ea

After substituting for $H_{1,2}$ using the dispersion relations, one has a five dimensional integral.  The structure functions $g_{1,2}(\nu,Q^2)$ depend only on two variables, and three integrals can be done analytically.  The results are
\ba
\label{eq:analytic}
\int \frac{ d^4 Q }{ Q^4 } \ \frac{ 2 Q^2 + Q_0^2 } { \tau Q^2 + 
Q_0^2 } \ldots 
&=& \pi^2 \int_0^\infty \frac{ d Q^2 }{ Q^2 }
\frac{1}{\tau} \, \beta_1(\tau)  \ldots 
\nonumber	\\[1.25ex] 
\int
\frac{ d^4 Q }{ Q^4 } \ \frac{ Q_0^2 } { \tau Q^2 +  Q_0^2 } 
\ldots &=& \pi^2 \int_0^\infty \frac{ d Q^2 }{ Q^2 } \,
\beta_2(\tau)  \ldots \,, 
\ea
where the ellipses stand for functions that may depend on $Q^2$ and other variables but not on the individual components of $Q$.   The auxiliary functions~\cite{Drell:1966kk,DeRafael:1971mc} are
\begin{eqnarray} 
\beta_1(\tau) &=& -3\tau + 2\tau^2 + 2(2-\tau)\sqrt{\tau(\tau+1)}  \,, 
		\nonumber \\ 
\beta_2(\tau) &=& 1 + 2\tau - 2 \sqrt{\tau(\tau+1)}		\,,
\end{eqnarray}
and have limits
\begin{eqnarray}
\beta_1(\tau) &=& \left\{
        \begin{array} {cl}
        4 \sqrt{\tau} + {\cal O}(\tau)	 \,, & \ \tau \to 0 ,  \\[1.25ex]
        \frac{9}{4} + {\cal O}(1/\tau) 
        %\left( 1 - \frac{5}{18} \frac{1}{\tau} + \frac{7}{48} \frac{1}{\tau^2} + \ldots \right)
	\,,	&      \  \tau\to\infty	,
        \end{array}
        \right.
                    \nonumber 	\\[1.5ex]
\beta_2(\tau) &=& \left\{
        \begin{array} {cl}
        1 + {\cal O}(\sqrt{\tau})\,, & \ \tau \to 0  ,   \\[1.25ex]
        0 + {\cal O}(1/\tau)
        %\frac{1}{4} \frac{1}{\tau}  - \frac{1}{8} \frac{1}{\tau^2} + \ldots 
        \,,	&      \  \tau\to\infty		.
        \end{array}
        \right.					
\end{eqnarray}

When the lepton mass can be neglected inside the integrals,  the results~(\ref{eq:analytic}) are directly applicable with $\tau = \nu^2/Q^2$.   When the lepton mass cannot be neglected, one has denominators that split into partial fractions,
\begin{align}
& \frac{1 }  { (Q^4 + 4m_\ell^2 Q_0^2) (Q^4 + 4m_p^2 Q_0^2) } 
			\nonumber	\\
& \quad = \frac{1}{4 Q^4 (m_p^2 -m_ \ell ^2)} \left( \frac{1}{ \tau_p Q^2 + Q_0^2}
	- \frac{1}{ \tau_ \ell Q^2 + Q_0^2 }	\right)	\,,
			\nonumber	\\ 
& \frac{1 }  { (Q^4 + 4m_ \ell ^2 Q_0^2) ( {\nu}^2 + Q_0^2) } 
			\nonumber	\\
& \quad = \frac{1}{ Q^2 (Q^2 - 4 m_ \ell ^2 \tau)} \left( \frac{1}{\tau Q^2 + Q_0^2 } 
	- \frac{1}{ \tau_ \ell Q^2 + Q_0^2}	\right)	\,.
\end{align}
Hence the results including lepton mass are given in terms of the same auxiliary functions and the notations
\be 
\tau_p \equiv \frac{Q^2}{4m_p^2} \,,	
\quad \tau_\ell  \equiv \frac{Q^2}{4m_\ell^2} \,,	
\ee
are useful.

The contribution of the two-photon diagram to the HFS becomes

\begin{widetext}

\begin{align}
& \frac{ E^{box}_{2\gamma} }{ E_F }    =  
\frac{ \alpha m_\ell m_p} { 2(1+\kappa_p) \pi (m_p^2-m_\ell^2) }
	\Bigg[ \int_0^\infty \frac{ d Q^2 }{ Q^2 } \ \left(    
\frac{\beta_1(\tau_p)}{\tau_p} - \frac{\beta_1(\tau_\ell)}{\tau_\ell}
\right) F_1(Q^2) G_M(Q^2)
			\nonumber	\\
&	+ 3 \int \frac{ d Q^2 }{ Q^2 } \   \Big( \beta_2(\tau_p) -
\beta_2(\tau_\ell)	\Big) \ F_2(Q^2) G_M(Q^2)	\Bigg]
+ \ \frac{ 2\alpha m_\ell} { (1+\kappa_p) \pi}
	\Bigg[ \int \frac{ d Q^2 }{ Q^2 } \int_{\nu_{th}}^\infty 
\frac{d\nu}{Q^2 - 4 m_\ell^2 \tau } \left(    
\frac{\beta_1(\tau)}{\tau} - \frac{\beta_1(\tau_\ell)}{\tau_\ell} \right)
g_1(\nu,Q^2)
			\nonumber \\
& - 3 \int \frac{ d Q^2 }{ Q^2 } 
\int_{\nu_{th}}^\infty  \frac{d\nu}{Q^2 - 4 m_\ell^2 \tau }  
\  \frac{Q^2}{ \nu^2} \Big( \beta_2(\tau) - \beta_2(\tau_\ell) \Big)  
\  g_2(\nu,Q^2)  \Bigg] \,.
\label{eq:boxstructure}
\end{align}

\end{widetext}

Conventionally, the proton structure-dependent corrections are split into three terms,
\be
\Delta_S = \Delta_Z + \Delta_R + \Delta_{\rm pol} = \frac{ E^{box}_{2\gamma} }{ E_F } - \frac{8\alpha m_r}{\pi}    \int_0^\infty \frac{dQ}{Q^2}	\,,
\ee
called Zemach, recoil, and polarizability terms.

We have subtracted from the box diagram the iteration of the lowest order one-photon exchange diagram, since in a bound state calculation that contribution is already included~\cite{Martynenko:2004bt}.  This  cancels the infrared divergence in the box diagram.  The visible effect of the subtraction is to give the ``$-1$'' in the Zemach term displayed below.   Including this subtraction gives the final result, up to a reorganization of terms.

The Zemach term is
\begin{align}
\label{eq:zemach} 
\Delta_Z &=  \frac{8 \alpha
m_r}{\pi} \int_0^\infty \frac{dQ}{Q^2} \left(
\frac{ G_E(Q^2) G_M(Q^2)}{1+\kappa_p} -1 \right) 
	\nonumber	\\	&\equiv  -2 \alpha m_r r_Z	\,,
\end{align}
where $r_Z$ is the Zemach radius.  The first part of the Zemach term is obtained from the first line of Eq.~(\ref{eq:boxstructure}) with $\beta_1(\tau_i)$ replaced by the first term in its low argument limit and $F_1$ replaced by $G_E$.  The Zemach term is finite in the nonrelativistic limit, meaning the proton mass going to infinity with electron mass and proton size held fixed.

Next continuing to the $g_1$ term, notice that aside from the overall $m_\ell$ factor, the integral diverges in the $m_\ell \to 0$ limit.  To demonstrate this, note that the $d\nu$ integral for $Q^2 =0$ in this limit is just a numerical factor times the left hand side of the Gerasimov-Drell-Hearn-Hosada-Yamamoto sum rule~\cite{Gerasimov:1965et,Drell:1966jv,hosoda1966},
\be 
\label{eq:gdh} 
4 m_p \int_{\nu_{th}}^\infty  \frac{d\nu}{\nu^2} 
	\    g_1(\nu,0) = - \kappa_p^2   \,.
\ee
The $dQ^2$ integral then diverges at its lower end.  

Conventionally, this near divergence is removed by adding an extra term to the $g_1$ integral.  The $g_1$ and $g_2$ terms together go into the polarizability correction, and with standard notation,
\be
\Delta_{\rm pol} = \frac{ \alpha m_\ell } { 2(1+\kappa_p) \pi m_p
} \left( \Delta_1 + \Delta_2 \right)	\,, 
\ee
with
\begin{align}
\Delta_1 &= \int_0^\infty	 
\frac{ d Q^2 }{Q^2 } 
\Bigg\{ \beta_1(\tau_\ell) F_2^2(Q^2) 
			\nonumber \\  
&+ \ 4 m_p
\int_{\nu_{th}}^\infty  \frac{d\nu}{\nu^2 } \ \frac{Q^4
\beta_1(\tau) - 4 m_\ell^2 \nu^2 \beta_1(\tau_\ell)}
{Q^4 - 4 m_\ell^2 \nu^2} g_1(\nu,Q^2)	
\Bigg\} 				,
\nonumber	\\[1.25ex] 
\Delta_2 &=  -12
m_p^2 \int_0^\infty	 \frac{ d Q^2 }{ Q^2 } 
			\nonumber	\\[1ex]
& \quad  \times  \int_{\nu_{th}}^\infty   \frac{d\nu}{\nu^2 }	
\ \frac{Q^4 \left( \beta_2(\tau) -
\beta_2(\tau_\ell) \right) }{Q^4 - 4 m_\ell^2 \nu^2} \
g_2(\nu,Q^2)					\,.
\end{align}

\noindent With the $F_2^2$ term, $\Delta_1$ is finite in any limit.   The reason for multiplying $F_2^2$ by $\beta_1(\tau_\ell)$ is to make it compatible with~\cite{Bodwin:1987mj,Pachucki:1996zza}.  Other choices could be made, including just using the numerical factor (9/4) as in the purely electron case~\cite{Drell:1966kk,DeRafael:1971mc}, but every choice has consequences elsewhere.

What remains is the recoil term,
\begin{align}
 \label{eq:recoil}
& \Delta_R^p =  \frac{2 \alpha m_r}{\pi m_p^2} 
	\int_0^\infty dQ \,
	F_2(Q^2) \frac{G_M(Q^2)}{1+\kappa_p}
					\nonumber	\\[1.25ex]
& + \ \  \frac{ \alpha m_\ell m_p} { 2(1+\kappa_p) \pi (m_p^2-m_\ell^2) }
	\Bigg\{
	\int_0^\infty    \frac{ d Q^2 }{ Q^2 }
					\nonumber	\\[1.25ex]
&	\times	\left(     \frac{\beta_1(\tau_p)-4\sqrt{\tau_p}}{\tau_p} 
		- \frac{\beta_1(\tau_\ell)-4\sqrt{\tau_\ell}}{\tau_\ell}		\right)
	F_1(Q^2) G_M(Q^2)  
					\nonumber	\\[1.25ex]
&	+ \ 3 \int_0^\infty  \frac{ d Q^2 }{ Q^2 } \   
		\Big( \beta_2(\tau_p) - \beta_2(\tau_\ell)	\Big) \ 
	F_2(Q^2) G_M(Q^2)	\Bigg\}
					\nonumber	\\[1.25ex]
& - \ \ \frac{ \alpha m_\ell } { 2(1+\kappa_p) \pi m_p }
	\int_0^\infty	 \frac{ d Q^2 }{ Q^2 } \ \beta_1(\tau_\ell) F_2^2(Q^2)		\,.
 \end{align}
The first term comes because the Zemach correction was written in terms of $G_E$ instead of $F_1$.  The middle terms can be recognized in Eq.~(\ref{eq:boxstructure}).  The last term compensates the extra term in the polarizability correction.

\begin{figure}[b]
\centerline{ \includegraphics[width=3.3in] {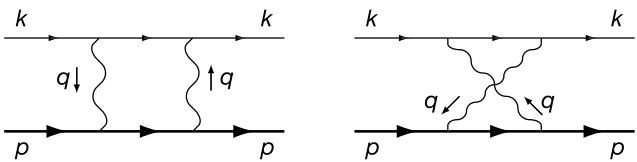} }
\caption{Elastic box and crossed  box.}
\label{fig:elastichfs} 
\end{figure}

Let us now recall a well-known albeit old-fashioned way to calculate the elastic contributions alone.  That is to calculate the box and crossed box diagrams, Fig.~\ref{fig:elastichfs}, using a photon-proton vertex given by
\be 
\label{eq:vertex} 
\Gamma_\nu(q) = \gamma_\nu F_1(Q^2) 
	+ \frac{i}{2m_p} \sigma_{\nu\rho}q^\rho F_2(Q^2)
\ee
for incoming photon momentum $q$.  We do not advocate doing the calculation this way because the above vertex cannot be complete or correct when the intermediate proton is off-shell.  One can correctly obtain the imaginary part of the elastic part of the Compton amplitude this way, since only on-shell configurations give the imaginary part.  However, this calculation has been done for the full box, and the results are well-known~\cite{Bodwin:1987mj,Pachucki:1996zza} and widely used (\textit{e.g.}~\cite{Dupays:2003zz,Martynenko:2004bt}) for the elastic terms.

%%%%%%%%%%%%%%%
%%%%%%%%%%%%%%%

The dispersive calculation, barring questions of subtractions, is complete and correct, and using the $F_2^2$ term described above ensures that the elastic terms---defined as the Zemach plus recoil terms---are the same as the result of calculating the box diagrams directly with the vertex of Eq.~(\ref{eq:vertex}).  To state the same thing in other words, if one uses the elastic contributions to the HFS found in~\cite{Bodwin:1987mj,Pachucki:1996zza} one must use the polarizability terms given here in order to obtain the correct total proton structure contributions to the HFS.

%%%%%%%%%%%%%%%%%%%%%%%%%%%%%%%%%%%%%%%%%%%%

\section{Convention Conversion Corrections}

\label{sec:correction}

%%%%%%%%%%%%%%%%%%%%%%%%%%%%%%%%%%%%%%%%%%%%

As noted, the $m_\ell \ne 0$ result for the polarizability correction in~\cite{Faustov:2001pn} differs from ours.  The difference is entirely in the $F_2^2$ terms.  They have
\be
\Delta_1 = 
	\int_0^\infty	 \frac{ d Q^2 }{ Q^2 } \Bigg[  \frac{9}{4}
	\beta_0(\tau_\ell) F_2^2(Q^2)
	 \hskip 0 em + ({\rm other\ terms\ same})	\Bigg],
\ee	 
 
\noindent with
\ba
\beta_0(\tau) = 2\sqrt{\tau(\tau+1)} - 2\tau .
\ea
This result is correct if used to complement the elastic terms calculated in the conventions of, for example,~\cite{Zinovev1970}.

We can, however, find two examples in the literature where the Ref.~\cite{Faustov:2001pn} polarizability corrections are combined with the elastic terms of Refs.~\cite{Bodwin:1987mj,Pachucki:1996zza} (the $\Delta_Z + \Delta_R^p$ as displayed here)~\cite{Dupays:2003zz,Martynenko:2004bt}.  In this case one should add a further correction for muonic hydrogen HFS given by
\ba
\Delta_{\rm pol}({\rm corr.}) &=&
\frac{\alpha m_r}{2(1+\kappa_p)\pi m_p} 
			\\ \nonumber
&\times&	\int_0^\infty	 \frac{ d Q^2 }{ Q^2 } \bigg[ \beta_1(\tau_\ell)-  \frac{9}{4} \beta_0(\tau_\ell) \bigg]
F_2^2(Q^2)  \,,
\ea
which evaluates to $-46$ ppm, or about $-1.0\ \mu$eV.  To two figures (in the ppm number), the result is the same using the form factors from Arrington, Melnichouk, and Tjon~\cite{Arrington:2007ux}, Kelly~\cite{Kelly:2004hm}, Arrington and Sick~\cite{Arrington:2006hm}, or the new Mainz form factors~\cite{Bernauer:2010wm,Vanderhaeghen:2010nd,Distler:2010zq}.

%%%%%%%%%%%%%%%%%%%%%%%%%%%%%%%%%%%%%%%%%%%%

\section{Effect upon the Lamb shift and proton radius}

\label{sec:lamb}

%%%%%%%%%%%%%%%%%%%%%%%%%%%%%%%%%%%%%%%%%%%%

In the context of the Lamb shift, the proton structure correction to the HFS that we discuss in this paper is small compared to the larger proton structure correction of the $2S$ energy level and has only a slight effect on the extracted proton radius.

The muon Lamb shift experiment measures the transition between the $2S_{1/2}^{F=1}$ and $2P_{3/2}^{F=2}$ states, and following the discussion in the introduction one may write an equation from Ref.~\cite{Pohl:2010zz} as
\begin{align}
& E(2P_{3/2}^{F=2}) - E(2S_{1/2}^{F=1}) = (r_p {\rm\ independent\ terms})
\nonumber \\ & \quad
 - 5.2262 {\rm\ meV}  \frac{r_p^2}{{\rm fm}^2}
+ 0.0347 {\rm\ meV}  \frac{r_p^3}{{\rm fm}^3} - \frac{1}{4} E_{HFS}^{2S}
\end{align}
where $r_p$ is the proton charge radius and the ``$r_p$ independent terms'' are well calculated.

A simple $46$ ppm reduction in the $2S$-state HFS results in an increase in $r_p$,  but one that is smaller than the uncertainty limit of the experiment~\cite{Pohl:2010zz}.  Stated differently, if one wants to make the above equality work with the CODATA value for the radius, one needs to find a $310 \ \mu$eV or so extra contribution to the energy, which we do not have from this source.

%%%%%%%%%%%%%%%%%%%%%%%%%%%%%%%%%%%%%%%%%%%%%%%%

\section{Closing Remarks}

\label{sec:end}

%%%%%%%%%%%%%%%%%%%%%%%%%%%%%%%%%%%%%%%%%%%%%%%%

We have reconsidered the hyperfine splitting for muonic hydrogen, advocating a unified calculation in order to clearly track how each term is defined.  We reported numerical results earlier, and have here detailed the derivation and shown an additional term that should be included if the ${\mathcal O}(\alpha^5)$ elastic and polarizability terms are taken from specific different publications.

The result of this term is to reduce the quoted hyperfine splitting by about $1.0 \ \mu$eV.

As a remark regarding the totality of the proton structure dependent corrections, many modern form factor fits give a Zemach radius toward the larger end of its former range~\cite{Brodsky:2004ck,Friar:2005rs,Brodsky:2005gz,Martynenko:2004bt}, and with a smaller spread~\cite{note1}.  Since the spread of values for the Zemach radius contributed the largest uncertainty to the calculated HFS, there could be a notable reduction in the quoted error limit.  The first three form factors listed in Table~\ref{table:hfs2s} show this possibility.  Table~\ref{table:hfs2s} was prepared using our own results for the ${\mathcal O}(\alpha^5)$ (lowest non-trivial order) proton structure corrections and using the extensive compilation of~\cite{Martynenko:2004bt} for the QED, higher-order, and additional small corrections, leaving these other terms in~\cite{Martynenko:2004bt} untouched and quoting results for the $2S$-state for definiteness.
We have also indicated the Zemach radius that follows from each form factor parameterization.  

\begin{table}[htdp]
\caption{Hyperfine splitting for the $2S$ state of muonic hydrogen, using different modern analytic fits in the terms that involve elastic form factors.}
\begin{center}
\begin{ruledtabular}
\begin{tabular}{ccc}
Form factor fit  & $E_{HFS}^{2S}$ (meV) &  $r_Z$ (fm)  \\ 
AMT~\cite{Arrington:2007ux}		& $22.8123$	& $1.080$		\\
Kelly~\cite{Kelly:2004hm}			& $22.8141$	& $1.069$		\\
AS~\cite{Arrington:2006hm}		& $22.8105$	& $1.091$		\\
Mainz 2010~\cite{Bernauer:2010wm,Vanderhaeghen:2010nd,Distler:2010zq}	
							& $22.8187$ &	$1.045$	
\end{tabular}
\end{ruledtabular}					
\end{center}
\label{table:hfs2s}
%\vglue - 3 mm
\end{table}%

However,  the latest electron scattering Mainz results~\cite{Bernauer:2010wm} create an exception.  They agree well with the CODATA value~\cite{Mohr:2008fa} for the charge radius but give a magnetic radius that is noticeably smaller than previous, but still modern, fits.  Along with this, they give a smaller Zemach radius~\cite{Vanderhaeghen:2010nd,Distler:2010zq} than the other form factor fits in Table~\ref{table:hfs2s}, and increase the spread of calculated values for $E_{HFS}^{2S}$.

Nearly all the uncertainty in the HFS calculation comes from the proton structure terms.  Quoting from~\cite{Carlson:2008ke}, the uncertainty in the polarizability contribution is $\pm\, 114$ ppm or $\pm\, 2.6 \ \mu$eV.    Determining the uncertainty in the elastic contributions from the spread of results from the selection of form factor parameterizations gives $\pm\, 4.1 \ \mu$eV 
($\pm\, 180$ ppm).  This would have been $\pm\, 1.8\ \mu$eV ($\pm 80$ ppm) before the latest result.  Adding in uncertainties in quadrature gives $\pm 4.9 \ \mu$eV, meaning that one should be allowed already to quote a smaller uncertainty as in 
\be
E_{HFS}^{2S} = 22.8146 (49) \rm{\  meV}\,,
\ee
compared with $E_{HFS}^{2S} = 22.8148 (78)$ meV obtained in~\cite{Martynenko:2004bt} and used in~\cite{Pohl:2010zz}.  One may expect that newer comprehensive analyses of the magnetic form factor data could reduce the uncertainty limit farther.  The central value quoted is the midpoint of the values in Table~\ref{table:hfs2s} and is little moved.   

This note has focused on one correction where we believed there were some definite statements to be made, and in the light of the PSI experiment on the Lamb shift,  all discrepancies need to be sorted out.   We have not reassessed the whole set of corrections to the HFS~\cite{Martynenko:2004bt,Borie:2004fv}, as has been done recently for the spin-independent case~\cite{Jentschura:2010ej,Jentschura:2010ha}.

\begin{acknowledgments} 
We thank Savely Karshenboim  and Zein-Eddine Meziani for useful conversations.  CEC thanks the National Science Foundation for support under grant PHY-0855618 and KG thanks the Department of Energy for support under grant DE-FG02-96ER41003.
\end{acknowledgments}

\bibliography{hfs}

\end{document}